\documentclass[12pt]{iopart}

\usepackage{etoolbox}
\apptocmd{\thebibliography}{\raggedright}{}{}

\usepackage{amsbsy}
\usepackage{amsfonts}
\usepackage{amsgen}
\usepackage{amssymb}
\usepackage[mathletters]{ucs} 
\DeclareUnicodeCharacter{183}{\ensuremath{\cdot}} 
\DeclareUnicodeCharacter{8469}{\ensuremath{\mathbb{N}}} 
\DeclareUnicodeCharacter{8474}{\ensuremath{\mathbb{Q}}} 
\DeclareUnicodeCharacter{8477}{\ensuremath{\mathbb{R}}} 
\DeclareUnicodeCharacter{8484}{\ensuremath{\mathbb{Z}}} 

\usepackage{listings}
\usepackage[utf8x]{inputenc}
\usepackage{color}
\usepackage[extendedchars,space]{grffile} 
\graphicspath{{./}{./resources/}}

\usepackage{colortbl}
\usepackage{slashbox}
\usepackage{array}
\usepackage{dcolumn} 

\usepackage{float}
\usepackage{todonotes}

\usepackage[english]{babel} \usepackage[english,num]{isodate}

\begin{document}

\title[Self-organized adaptive stochastic tipping]{Self-organized
       stochastic tipping in slow-fast dynamical systems}

\author{Mathias~Linkerhand, Claudius~Gros}
\address{Institute for Theoretical Physics, Goethe University,
         Frankfurt a.\,M., Germany}
\ead{linkerhand[[at]]th.physik.uni-frankfurt.de}
\ead{gros[[at]]th.physik.uni-frankfurt.de}

\date{2012-02-23} 

\begin{abstract}
Polyhomeostatic adaption occurs when evolving systems
try to achieve a target distribution function
for certain dynamical parameters, a generalization
of the notion of homeostasis. Here we consider a
single rate encoding leaky integrator neuron model
driven by white noise, adapting slowly its internal 
parameters, the threshold and the gain, in order to achieve 
a given target distribution for its time-average firing rate. For the case of sparse encoding, when
the target firing-rated distribution is bimodal, we
observe the occurrence of spontaneous quasi-periodic
adaptive oscillations resulting from fast transition
between two quasi-stationary attractors. We interpret
this behavior as self-organized stochastic tipping, with
noise driving the escape from the quasi-stationary attractors.
\end{abstract}

\maketitle
\section{Introduction}
Self-regulation plays an important role in biological and technical systems.
Homeostatically regulated steady states are a precondition to life, examples
being the concentration of blood glucose controlled by insulin \cite{plum06} and
glucagon, the pH value of blood \cite{schaefer61,tresguerres10} and the body
temperature \cite{charkoudian03}, which are all autoregulated in order to
maintain stable conditions. Further examples are the concentration of ions,
proteins and transmitters in the brain, their respective levels are all self
regulated \cite{marder06}. Furthermore, homeostasis is implemented and can be
found in technical systems, for example in microrobotic swarms
\cite{kernbach11}.
Adaption typically introduces a slow time scale into the dynamical system
\cite{grosBook}, a process also denoted meta learning, a central notion in the
context of neuromodulation \cite{doya02} and emotional control \cite{gros10}.
The resulting dynamical system then has both fast and slow variables and
critical transitions in the form of tipping processes may occur \cite{kuehn11}.

Classical homeostasis involves the regulation of a scalar quantity, like the
body temperature. More complex forms of homeostasis are however also important
in the realm of life. For example, an animal may want to achieve a certain time
averaged distribution of behaviors, like foraging, resting and engaging
socially, over the course of several days.
This kind of adaptive behavior has been termed polyhomeostasis
\cite{markovic10,markovic12}. It occurs when a dynamical system tries to
achieve, via the continuous adaption of slow variables, a given target
distribution for the time-averaged activity of a subset of fast variables.
Polyhomeostatically adapting systems are typically slow-fast dynamical systems
and their dynamical behavior can tip spontaneously from one state into another.
For a network of rate-encoding neurons tipping transitions from laminar flow to
intermittent chaotic bursts of activities have been observed
\cite{markovic10,markovic12}.

Tipping transitions can occur both in adaptive and in driven systems. 
Potential tipping scenarios are currently discussed intensively in the 
context of climate research \cite{ashwin12,lenton08}, they may be 
related to a slow driving of external parameters \cite{baer89}, to 
noisy input inducing a stochastic escape from a local attractor 
\cite{gammaitoni98,mcdonnell2009}, or through a dynamical effect 
when the rate of change of a control parameter reaches a certain
threshold \cite{ashwin12}.

Here we study the phenomenon of self-organized tipping for a 
polyhomeostatic adapting system driven by a steady-state stochastic 
input. We examine a previously proposed model 
\cite{triesch05a, stemmler99} for regulating the firing rate 
distribution of individual neurons based on information-theoretical
principles. This type of model has been studied previously for the 
case of discrete time systems and unimodal target firing rate 
distributions \cite{markovic10,markovic12}. Here we examine the 
case of continuous time and bimodal target distribution functions, 
corresponding to sparse coding. For bimodal firing rate distributions 
the neural activity tends to switch in between states close to minimal
and maximal activity. Similar bimodal activity states are observed 
also in many other systems, e.g.\ dynamical gene regulation 
networks \cite{davidson2006}. We find that bimodal target 
distributions may lead to self-organized 
bistability within a certain range of parameters.
 
We consider a single leaky integrator neuron with noisy input and 
a sigmoidal transfer function having two degrees of freedom. To 
achieve a special behavior -- here the temporal output distribution 
of the firing rate -- we use polyhomeostasis to change the intrinsic 
parameters which are directly influencing the mapping of the membrane 
potential to the firing rate in order to obtain a specific output 
distribution. We derive these parameter changing rules using 
stochastic adaption and show that two degrees of freedom already 
result in a good behavior approximation, for most of the parameters 
studied. For bimodal adaption target distributions we observe 
self-organized and quasi periodic stochastic tipping in between 
two quasi-stationary attractors resulting from competing adaption 
gradients.

\begin{figure}[t]
\centering
\includegraphics[width=0.6\columnwidth]{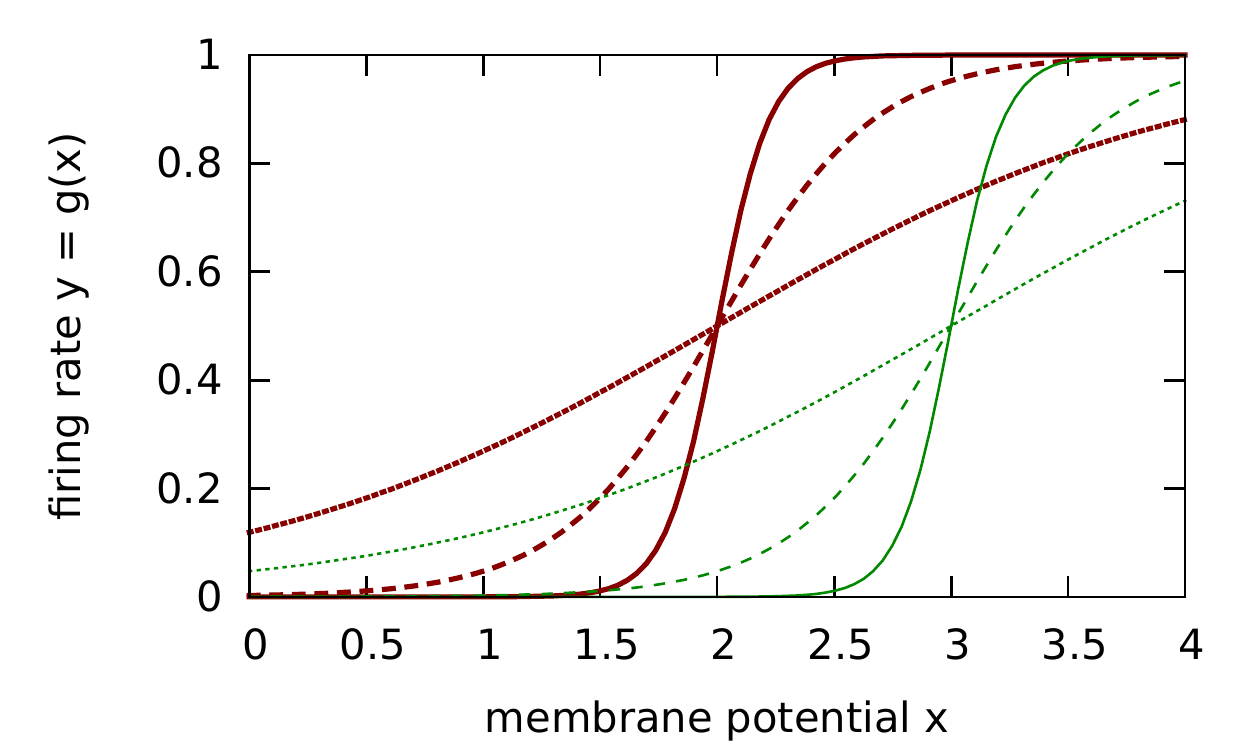}
\caption{The transfer function $g(x)$, see Eq.~(\ref{eq_g}),
for thresholds $b=2$ (red lines) and $b=3$ (green lines) and 
various gains $a$: 1 (dotted), 3 (dashed), 9 (solid).}
\label{fig_transferFunction}
\end{figure}

\section{Model}
Biological neurons integrate incoming signals and 
emit an axon potential, a spike, whenever the membrane
potential has reached a certain threshold. The membrane
potential then returns, after a short refractory period,
rapidly to its resting value. This 
behavior can be captured using spiking
integrate and fire neural models \cite{burkitt06}.
In many circumstances the firing rate, the number of
spikes per unit time, is important and rate encoding neural
models can be used \cite{borst99}. Here we consider a single 
rate-encoding leaky integrator
driven by white noise $\xi(t)$,
\begin{equation}
\dot x(t) = - \Gamma  x(t) + \xi(t),
\qquad
\langle\xi(t)\xi(t')\rangle=Q\delta(t-t')~,
\label{eq_model}
\end{equation}
where $x>0$ is the membrane potential and $\Gamma>0$
the relaxation rate. The firing rate $y(t)\in[0,1]$
is a nonlinear function of the membrane potential
$x(t)$, which we have selected as
\begin{equation}
y(t) = g(x(t)),
\qquad
g(x) = \frac{1}{1 + e^{-a (x - b)}}~,
\label{eq_g}
\end{equation}
where $a>0$ is the gain and 
$b$ is the threshold. The polynomial transfer
function (\ref{eq_g}) maps 
the membrane potential $x\in[-\infty,\infty]$
to the normalized firing rate $y\in[0,1]$
which approaches zero and unity for small and large
membrane potentials respectively, compare
Fig.~\ref{fig_transferFunction}. The slope of $g(x)$ is
$a/4$ at the threshold $b$.

Usually the intrinsic parameters of the transfer
function (\ref{eq_g}), $a$ and $b$ 
are taken as given by some a priori considerations.
Here we will consider them to be slow variables,
$a=a(t)$ and $b=b(t)$, adapting slowly such that
a target dynamical behavior is approached on the
average for the firing rate $y(t)$.
The stochastic driving $\xi (t) \in  [\Xi _1, \Xi _2]$ in (\ref{eq_model})
is simulated through white noise plateaus: The values are generated according to a uniform probability distribution (white noise), but they
remain constant for short time intervals on the
order of unity.
The membrane potential averages the input driving
noise, due to the leak rate $\Gamma$ in (\ref{eq_model}),
its distribution function $\rho(x)$ having a
mean $\mu_\rho \approx(\Xi_1+\Xi_2)/(2\Gamma)$ and variance
$\sigma_\rho^2 \approx(\Xi _2-\Xi_1)/(2\Gamma)$.


\subsection{Polyhomeostatic Adaption}

The firing-rate statistics is given by
\begin{equation}
p(z) = {1\over T}\int_{t_0}^{t_0+T}\!\delta(z-y(t))\,\mathrm dt,
\qquad
\int_0^1 p(z)\mathrm\,dz = 1~,
\label{eq_def_p}
\end{equation}
where the length $T$ of the sliding observation window is
substantially larger than the relaxation rate
$1/\Gamma$. The firing-rate distribution $p(z)$
is an important quantity characterizing the
information processing capability of biological
and artificial neurons. No information is encoded 
for a constant firing rate, the next value is always exactly the same as before, so no new information is transferred. One may assume that 
a certain distribution $q(y)$ of firing rates may 
constitute an optimal working regime. Possible
functional dependencies for $q(y)$ can be derived
by information-theoretical considerations, e.g.\ 
maximizing information entropy, as discussed further
below.

Considering a given target firing-rate distribution
$q(y)$, the closeness of the actual firing-rate distribution
$p(y)$ is measured by the Kullback-Leibler divergence 
(KL-divergence0 $D_\mathrm{KL}$, their relative entropy, 
\cite{grosBook}:
\begin{equation}
D_\mathrm{KL}(p, q) = \int \! \mathrm dy \, p(y) 
     \ln \frac{p(y)}{q(y)},
\qquad 
D_\mathrm{KL}(p, q) \ge 0~.
\label{eq_KL}
\end{equation}
The Kullback-Leibler divergence is positive definite and
vanishes only when the two distribution coincide. The
KL-divergence is generically not symmetric but becomes
symmetric in the limiting case of similar distributions
$p$ and $q$, becoming equivalent in this limit to 
the $\chi^2$ test \cite{grosBook}.
Our aim is now to rewrite (\ref{eq_KL}) as an integral 
over the membrane potential $x$, using
\begin{equation}
p(y) \mathrm dy = \rho (x) \mathrm dx,
\qquad 
p(y) = \frac{\rho (x)}{{\mathrm dy}/{\mathrm dx}}~,
\label{eq_x_y}
\end{equation}
where $\rho(x)$ is the membrane potential distribution.
Using $y=g(x)$ and Eqs.~(\ref{eq_KL}) and (\ref{eq_x_y}),
we obtain 
\begin{equation}
\frac{\partial D_\mathrm{KL}}{\partial \theta } = 
\int \! \mathrm dx\, \rho (x) \left[ -\frac{1}{g'} 
\frac{\partial g'}{\partial \theta } - 
\frac{q'}{q} \frac{\partial g}{\partial \theta } \right]
\equiv 
\int \! \mathrm dx\, \rho (x)\,
\frac{\partial d}{\partial \theta }
\label{eq_D_KL_theta}
\end{equation}
for the derivative of the Kullback-Leibler divergence with
respect to the intrinsic parameters $\theta=a,\, b$ 
of the transfer function $g(x)$, see (\ref{eq_g}).

We consider now the case that the system does not
dispose of prior information about the distribution
of input stimuli and the thereby resulting distribution
of membrane potential $\rho(x)$. The best strategy to
minimize the Kullback-Leibler is then to minimize
the individual terms of the integral (\ref{eq_D_KL_theta})
through the stochastic adaption rules
\cite{triesch05a,markovic10}
\begin{equation}
\frac{\mathrm{d} \theta}{\mathrm{d} t} = - \epsilon _\theta\,
\frac{\partial d}{\partial\theta},
\qquad
\theta = a,\ b
\label{eq_deltaTheta}
\end{equation}
for the intrinsic parameters of the transfer function 
$g(x)$, where the $\epsilon_\theta$ are appropriate small
adaption rates.

\begin{figure}[t]
\centering
\includegraphics[width=0.6\columnwidth]{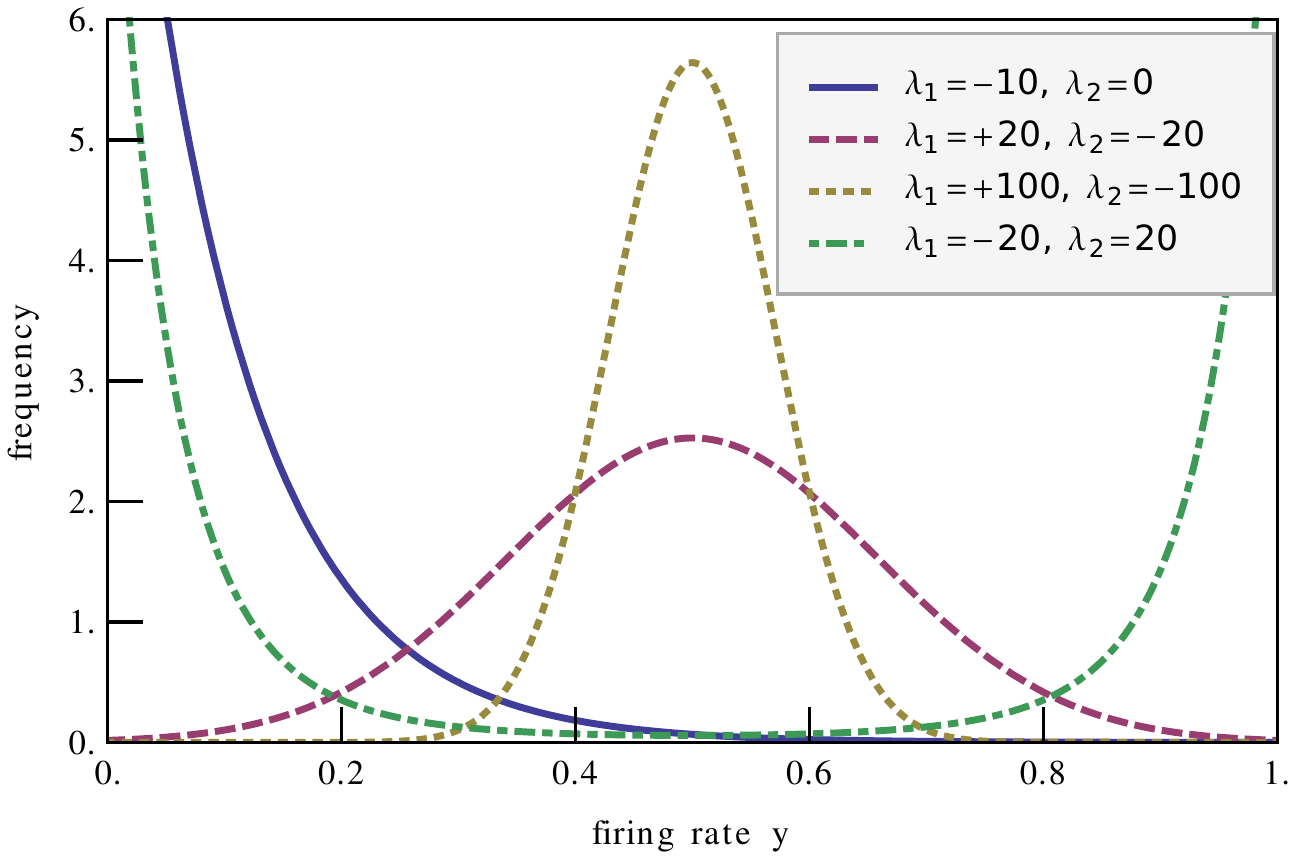}
\caption{Target distribution $q(y)$, see (\ref{eq_q_gaussian}),
with some selected parameters $\lambda _1$ and $\lambda _2$.
The target firing-rated distributions is bimodal for 
$\lambda_2 > 0$.}
\label{fig_targetDistributions}
\end{figure}

\subsection{Target Firing-Rate Distribution}
In order to evaluate (\ref{eq_deltaTheta}),
respectively Eq.~(\ref{eq_D_KL_theta}), we 
need to specify the target firing rate
distribution $q(y)$. For this purpose we use
information-theoretical considerations.

Given a continuous probability distribution 
function $q$ its Shannon entropy $H(q)$ can be defined as
\begin{equation}
H(q) = - \int \mathrm dy \, q(y) \ln q(y)
\end{equation}
Among all the real-valued distributions with 
specified mean $μ$ and standard deviation $σ$ 
the Gaussian distribution \cite{grosBook} 
\begin{equation}
q(y) \propto
\exp \left( -\frac{\left( y - μ \right)^2}{2 σ^2} \right)
\propto \exp\left(\lambda_1 y + \lambda_2 y^2\right)
\label{eq_q_gaussian}
\end{equation}
has maximal information entropy, with
$\mu=-\lambda_1/(2\lambda_2)$ and $2\sigma^2=-1/\lambda_2$, 
which is easily 
obtained using variational calculus,
\begin{equation*}
0 = \delta\left[
H(q) + \lambda_1\int \mathrm dy\, y\, q(y)
     + \lambda_2\int \mathrm dy\, y^2\, q(y)
\right]~,
\end{equation*}
where $(-\lambda_1)$ and $(-\lambda_2)$ are the 
respective Lagrange parameters.
In Fig.~\ref{fig_targetDistributions} examples for
$q(y)$ are illustrated for several values of
$\lambda_1$ and $\lambda_2$. The support of the
target firing rates is compact, $y\in[0,1]$,
and both negative and positive $\lambda_1$ 
and $\lambda_2$ can be considered.
The normalization factor
$\int_0^1 \mathrm dy\,q(y)$ cancels out in (\ref{eq_D_KL_theta}), since only ratios are involved.

For positive $\lambda_2 > 0$ and $\lambda_1 \approx -\lambda_2$
one obtains bimodal target distributions.
This is an interesting case, since sparse coding, which
is realized when only a minority of neurons of a given
network is active, and a majority is inactive
\cite{olshausen04}, is
characterized by a skewed bimodal distribution.

\begin{figure}[t]
\centering
\includegraphics[width=0.6\columnwidth]{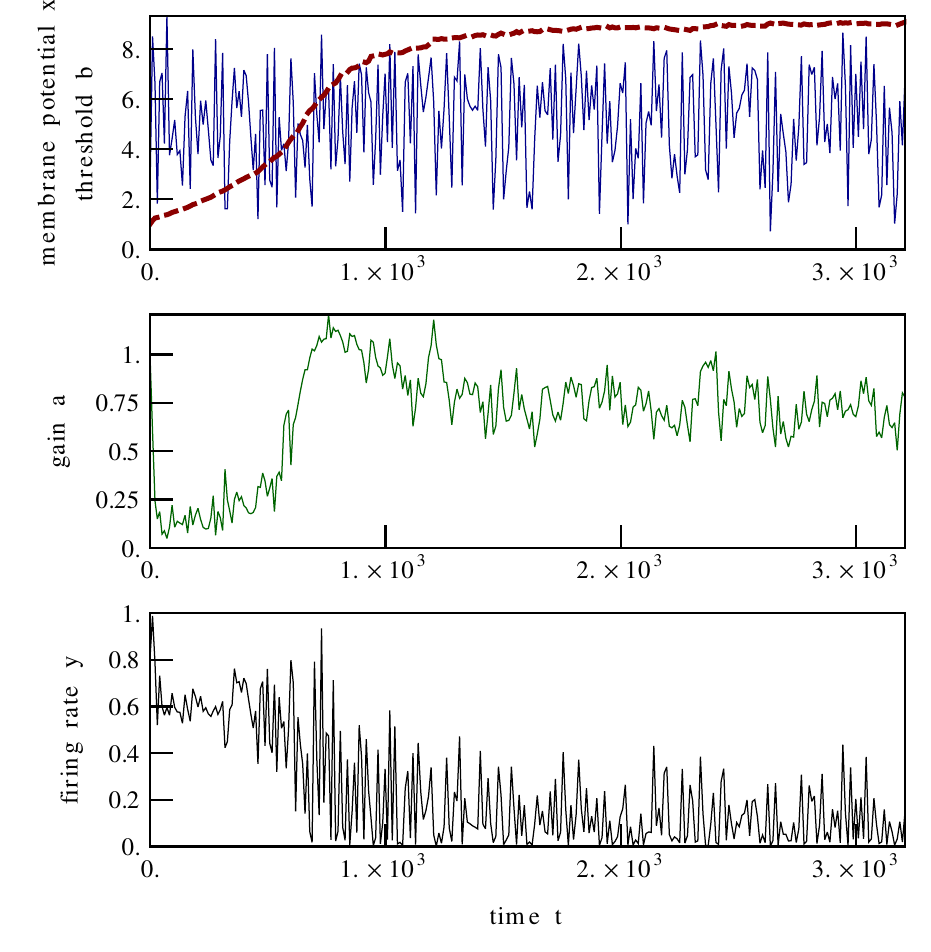}
\caption{Typical time series for a mono-modal target distribution
$q(y)$ with  $\lambda _1 = -10$, $\lambda _2 = 0$,
compare Fig.~\ref{fig_targetDistributions}. Plotted
are the membrane potential $x$ (solid blue line, upper panel), 
the threshold $b$ (dashed red line, upper panel), 
the gain $a$ (solid green line, middle panel)
and the firing rate $y$ (solid black line, lower panel). 
$\Delta t = 10^{-1}$, $\epsilon _a = \epsilon _b = 10^{-2}$, 
$\Gamma  = 1$. }
\label{fig_timeSeries}
\end{figure}

\subsection{Stochastic Adaption Rules}

From (\ref{eq_q_gaussian}) and (\ref{eq_g})
we find the relations
\begin{equation*}
\frac{q'(y)}{q(y)} = \lambda _1 + 2 \lambda _2 y,
\qquad
\frac{\partial g}{\partial x} = a g \left(1 - g\right)
\end{equation*}
and
\begin{equation*}
\frac{\partial g}{\partial a} = 
(x-b) g \left(1 - g\right),
\qquad
\frac{\partial g}{\partial b} = 
-a g \left(1 - g\right)~,
\end{equation*}
which we can use to evaluate the stochastic 
adaption rules (\ref{eq_deltaTheta}) as
\begin{equation}
\frac{\mathrm{d}a}{\mathrm{d}t} = \epsilon _a \left[ \frac{1}{a} + (x-b)
\Big[ 1 - 2 y + \left( \lambda _1 + 2 \lambda _2 y \right) 
\left( 1 - y \right) y \Big] \right]
\label{eq_delta_a}
\end{equation}
and
\begin{equation}
\frac{\mathrm{d}b}{\mathrm{d}t} = \epsilon _b \left[ - a
\Big( 1 - 2 y + \left( \lambda _1 + 2 \lambda _2 y \right) 
\left( 1 - y \right) y \Big) \right]~.
\label{eq_delta_b}
\end{equation}
These two adaption rules will lead to an
adaption of the time-averaged firing rate
distribution $p(y)$ towards the target
distribution $q(x)$ whenever the adaption 
time-scales $1/\epsilon_\theta$ are 
substantially larger than the time constants
of the neural dynamics, which in turn are
determined by the time scale of the incoming
stimuli and by the leak-rate $\Gamma$ in
(\ref{eq_model}).

The transfer function $g(x)$ contains only two free
parameters, the gain $a$ and the threshold $b$.
Perfect adaption $p(y)\equiv q(y)$, for all $y\in[0,1]$
can hence not be expected. The system tries to
minimize the Kullback-Leibler divergence by
adapting the available degrees of freedom, which
are just two in our case.

\begin{figure}[t]
\centering
\includegraphics[width=0.6\columnwidth]{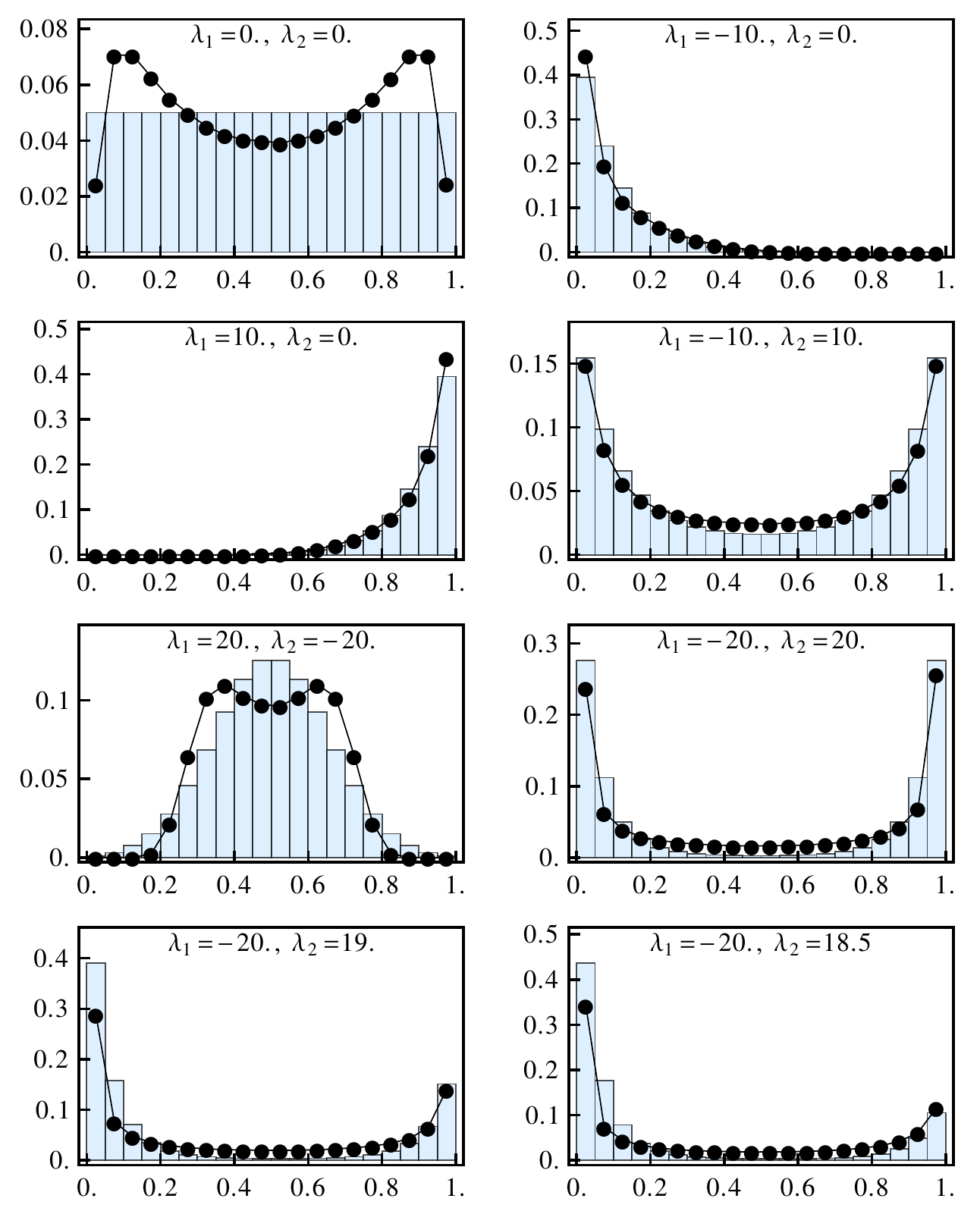}
\caption{Target distribution $q$ (bars) vs.\ achieved 
distribution $p$ (points) for different distributions. 
$\lambda_1$ and $\lambda_2$ are given in each diagram.
$\Delta t = 10^{-1}$, $t_\mathrm{max} = 10^8$,  
$\epsilon _a = \epsilon _b = 10^{-2}$, 
$\Gamma = 1$, $\Xi = [0, 10]$.}
\label{fig_targetVsAchievedDistributions}
\end{figure}

\subsection{Numerical method}
The equations (\ref{eq_model}), (\ref{eq_delta_a}) and (\ref{eq_delta_b}) form a set of first order differential equations with respect to time. We solve them numerically using the Euler method with one evaluation per time step. The random white noise is generated through a pseudo-random number generator with a uniform distribution. The values for the leak $\Gamma$, the time step $\Delta t$ and the learning rates $\epsilon_a$ and $\epsilon_b$ are shown in the corresponding figures.

\section{Results}
We performed a series of simulations with the aim
to study two issues. Polyhomeostatic adaption had
been studied previously for the case of discrete time systems
\cite{triesch05a,markovic10}, here we examine the
case of continuous time. The case of a bimodal target
distribution is, in addition, highly interesting,
as it confronts the system with a dilemma.
The transfer function $g(x)$, compare 
Fig.~\ref{fig_transferFunction}, is strictly monotonic.
The distribution of the membrane potential $\rho(x)$
is hence mono-modal. There is no easy way for the adapting
neuron to achieve, as a steady state time-average, 
a bimodal output firing rate distribution $p(y)$.
The question then is whether the system will find
a way out of this dilemma through spontaneous
behavioral changes.

\subsection{Target Distribution Approximation}

For most simulations we used, if not stated otherwise,
$\Gamma=1$ for the leak rate and $\Delta t=10^{-1}$ for
the integration time step. A typical time 
series is given in Fig.~\ref{fig_timeSeries}. Note that
the adaption of the intrinsic parameters $a$ and $b$
takes place on a slower time scale than the one of the
primary dynamic variables, $x$ and $y$, as typical for a
slow-fast dynamical system.

Applying moderate to small learning rates 
$\epsilon_a = \epsilon_b \lesssim 0.01$ 
the neuron’s firing rate $y$ approximates various types 
target distributions $q$ quite well. In 
Fig.~\ref{fig_targetVsAchievedDistributions} the achieved 
and the respective target firing rated distributions are compared.
The respective relative entropies are well minimized and 
presented in Table~\ref{tab:relativEntropiesVariousTargetDistributions}.
Strictly speaking the stochastic adaption rules 
(\ref{eq_delta_a}) and (\ref{eq_delta_b}) are equivalent to
approximating the firing-rate statistics (\ref{eq_def_p}), which
is a time-averaged quantity, towards the target distribution 
function $q(y)$ only in the limit of very small adaption rates,
$\epsilon_a$ and $\epsilon_b$. Small but finite values for the
adaption rates, as used in our simulations, correspond to
to a trailing averaging procedure over a limited time interval,
and the value of Kullback-Leibler divergence achieved hence depend
weakly on the actual values used for the learning rates.

For very high learning rates, $\epsilon_b \gg 0.1$, the threshold 
$b$ follows the membrane potential $x$ nearly instantaneously, 
both variables become highly correlated. Therefore the firing rate distribution 
$p$ cannot approximate the target distribution $q$ any more,
in fact the resulting Kullback-Leibler divergence is then very high.
The tipping in dynamic behavior as a function of adaption rate
amplitude is typical for a rate-induced tipping 
transition \cite{ashwin12}.

\begin{table}[b]
\caption{\label{tab:relativEntropiesVariousTargetDistributions}
The relative entropies $D_{KL}$ (\ref{eq_KL})
of various target distributions 
(see Fig.~\ref{fig_targetDistributions})
compared to the corresponding achieved distribution,
compare Fig.~\ref{fig_targetVsAchievedDistributions}.}
\begin{indented}
	\lineup
\item[]\begin{tabular}{llll}
	\br
	$\lambda _1$ & $\lambda _2$ & shape & $D_{KL}$ \tabularnewline
	\mr
	0 & 0 & uniform & 0.043 \tabularnewline
	-10 & 0 & left dominant & 0.034 \tabularnewline
	+10 & 0 & right dominant & 0.028 \tabularnewline
	-10 & +10 & left/right dominant & 0.018 \tabularnewline
	\mr
	+20 & -20 & hill & 0.076 \tabularnewline
	-20 & +20 & left/right, symmetric & 0.175 \tabularnewline
	-20 & +19 & left/right, left skewed & 0.244 \tabularnewline
	-20 & +18.5 & left/right, left skewed & 0.283 \tabularnewline
	\br
\end{tabular}
\end{indented}
\end{table}

\subsection{Gain-Threshold Phase Diagram}
\begin{figure}[t]
\centering
\includegraphics[width=0.6\columnwidth]{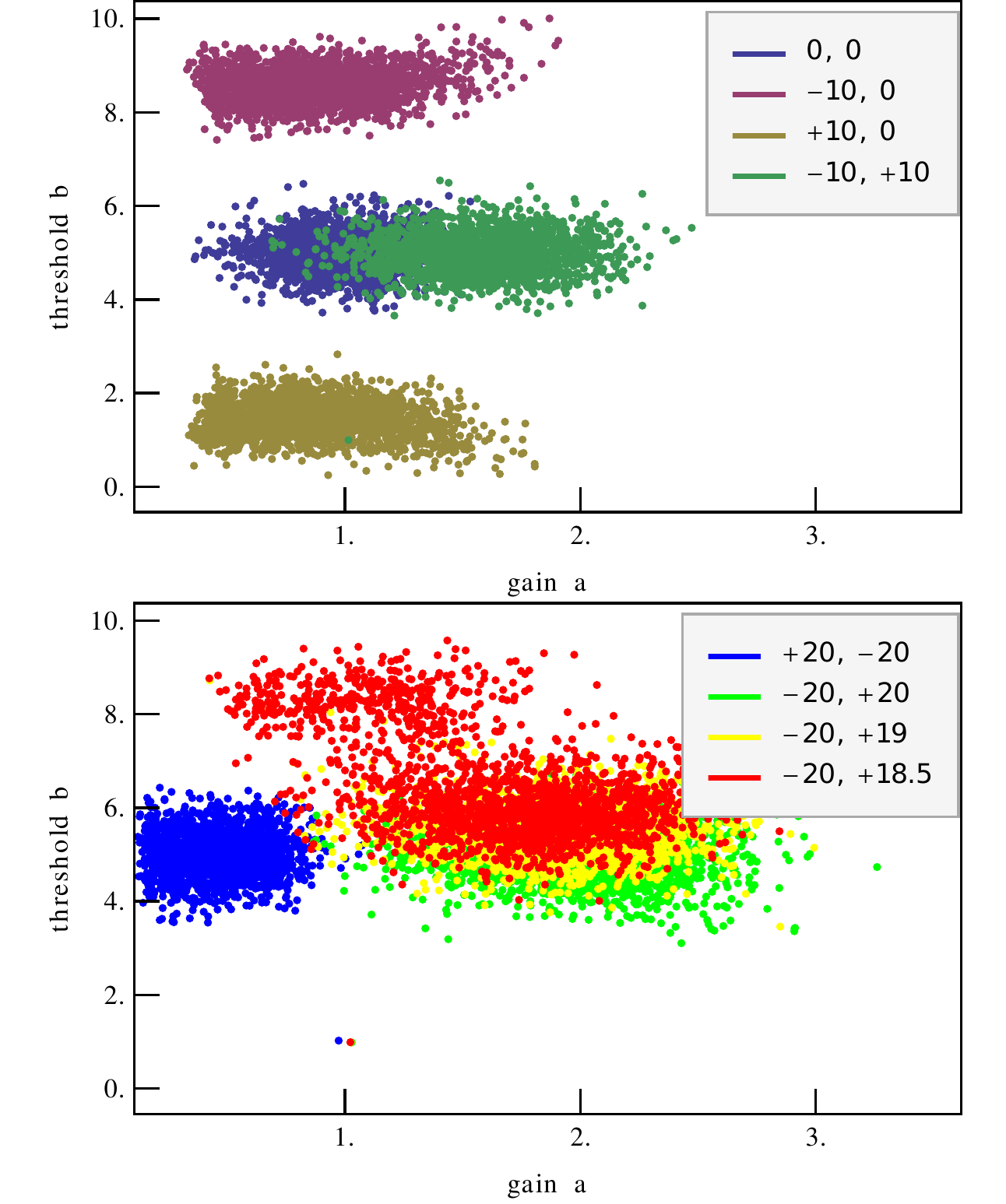}
\caption{Phase diagram. Plotted are the gain $a(t)$ and the
threshold $b(t)$ of the transfer function for various target 
distributions ($\lambda_1$ and $\lambda_2$ given in the legend). 
The respective target and achieved firing rate distributions
are given in Fig.~\ref{fig_targetVsAchievedDistributions}.
}
\label{fig_gainThresholdPhaseDiagramVariousTargetDistributions}
\end{figure}

Due to the sigmoidal shape of the transfer function, 
several target distributions lead to specific fingerprints 
in the gain-threshold phase diagram which we present in
Fig.~\ref{fig_gainThresholdPhaseDiagramVariousTargetDistributions}.
The threshold, for example, for a left (right) dominant 
target distribution is high (low) and is therefore sensitive 
to the mean $\mu=-\lambda_1/(2\lambda_2)$ of $q(y)$. 
Small gains $a$ result in quite flat transfer functions
$g(x)$, compare Fig.~\ref{fig_transferFunction}, mapping 
the membrane potentials to similar firing rates $y$.
High gains $a$ discriminate, relative to the threshold $b$,
on the other side between high and low membrane potentials.
The gain is therefore smaller for hill shaped and 
flat target distributions, as compared 
to the left and right dominant target distributions
(e.g.\ $\lambda_1 = -20$, $\lambda_2 = +20$) for which
intermediate values are suppressed.

Left (right) dominant target distributions 
(compare Table~\ref{tab:relativEntropiesVariousTargetDistributions}) 
correspond directly to high (low) transfer function thresholds. 
Uniform, hill and other not unilateral dominant target 
distributions lead to intermediate transfer function thresholds 
with a wide variety of the transfer function gains. 
For symmetrical target distributions from hill shaped 
to diametrical shaped there is a transition from low to high gains.

\begin{figure}
\centering
\includegraphics[width=0.8\columnwidth]{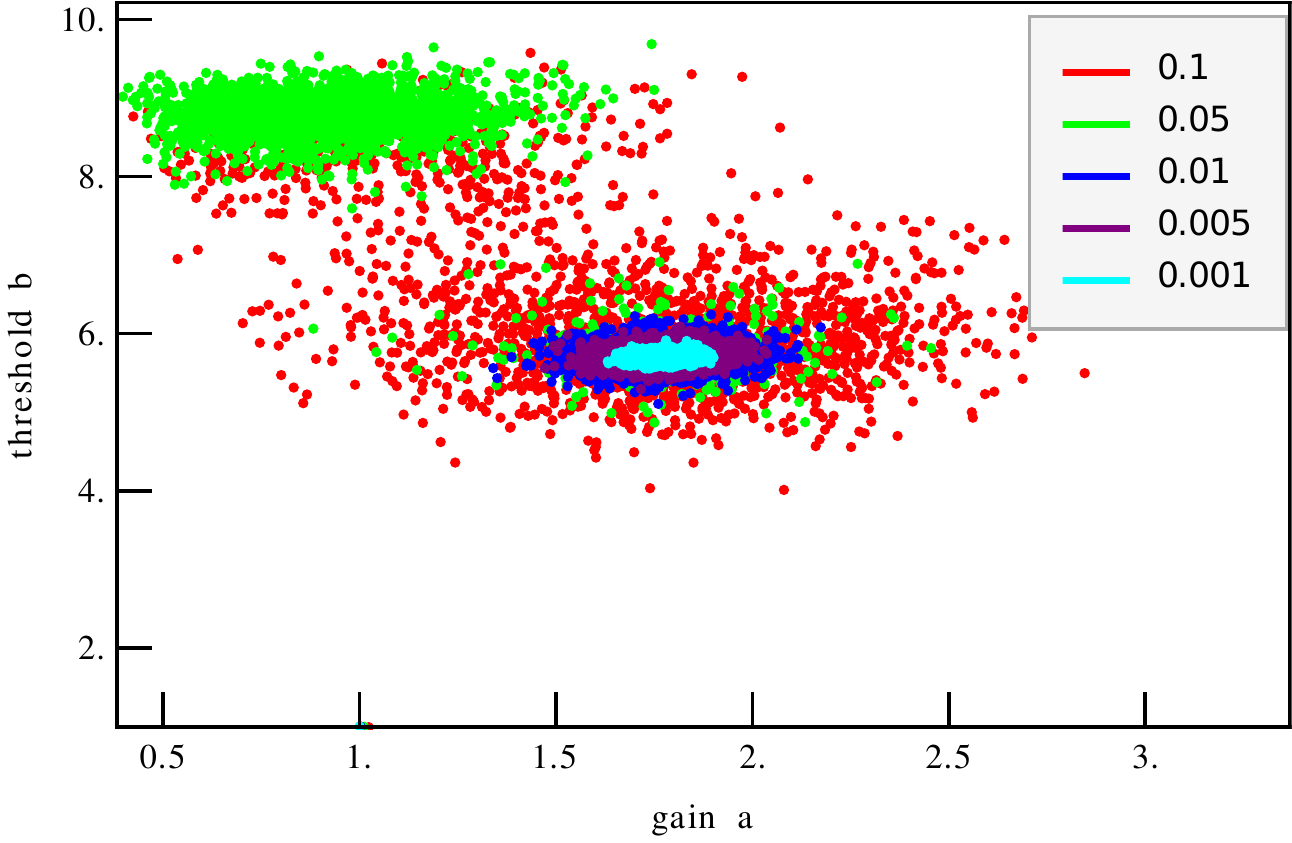}
\caption{Stochastic escape: Phase diagram of the transfer 
function gain vs.\ transfer function threshold for a convex 
left skewed target distribution with various learning 
rates ($\epsilon _a = \epsilon _b$ given in the legend). 
$\Delta t = 10^{-1}$, $\Gamma  = 1$, 
$\lambda _1 = -20$, $\lambda _2 = 18.5$.}
\label{fig_stochastic_escape}
\end{figure}

\subsection{Self-Organized Stochastic Escape}

While the left or right dominant target distributions 
are easily approximated due to the sigmoidal shape 
of the transfer function, the bimodal left and right dominant 
target distributions puts the system in dilemma: 
Since intermediate values are to be suppressed the 
transfer function gain $a$ cannot be too small. Because 
of this there exists at least two quasi-stationary fixed points, 
one for the left, one for the right part of the distribution.

For zero or small learnings rates $\epsilon_a = \epsilon_b \approx 0$ 
the system is trapped in a single local fixed point. Only 
the left or right part of the target distribution is 
then approximated, the Kullback-Leibler divergence 
is not well minimized.

\begin{table}[b]
\caption{\label{tab:relativEntropies}
Relative entropies $D_{KL}$ (\ref{eq_KL}) for the left-skewed target
distribution ($\lambda _1 = -20$, $\lambda _2 = 18.5$) relative to
the achieved distribution for various learning rates $\epsilon _a$ 
and $\epsilon _b$, compare 
Fig~\ref{fig_stochastic_escape}.}
\begin{indented}
        \lineup
\item[]\begin{tabular}{l|lllllll}
        \br
        $\epsilon _a = \epsilon _b$ & $10^{-5}$ & $10^{-4}$ & $10^{-3}$ & $5 \cdot 10^{-3}$ & $10^{-2}$ & $5 \cdot 10^{-2}$ & $10^{-1}$ \tabularnewline
        \mr
        $D_{KL}$ & 0.306 & 0.295 & 0.293 & 0.289 & 0.283 & 0.154 & 0.109 \tabularnewline
        \br
\end{tabular}
\end{indented}
\end{table}

\begin{figure}
\centering
\includegraphics[width=0.6\columnwidth]{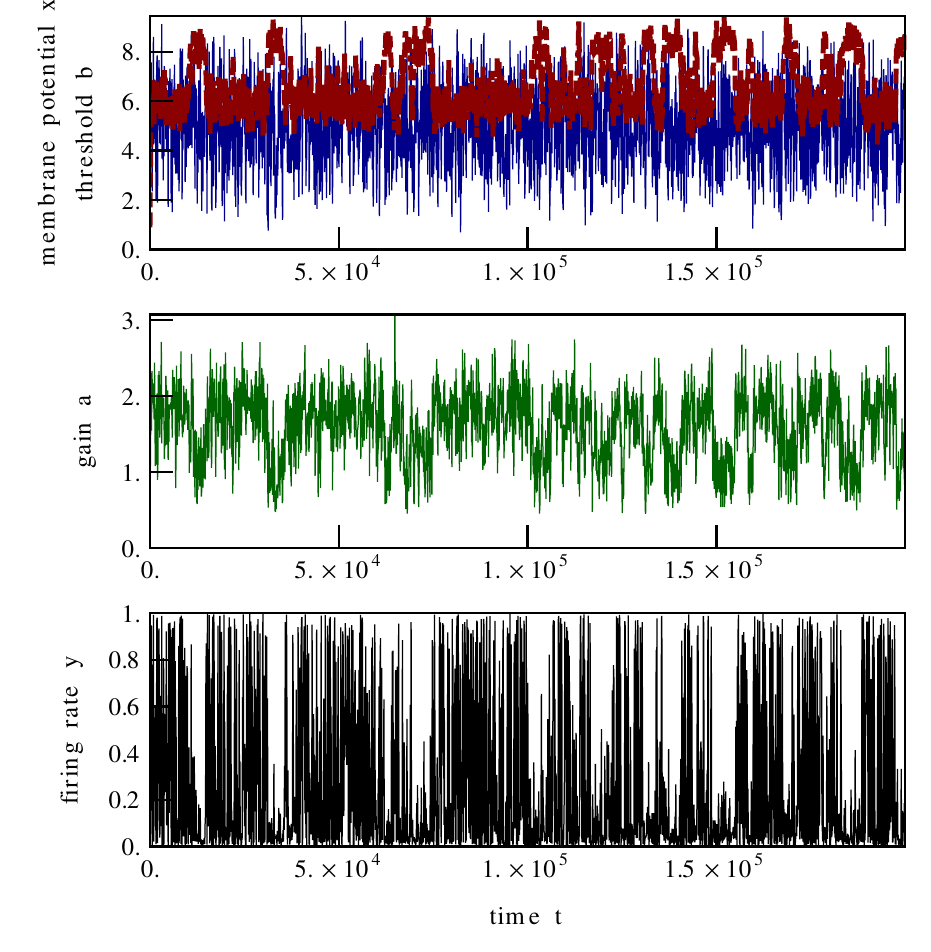}
\caption{Time series: membrane potential $x$, 
transfer function threshold $b$ (dashed), 
transfer function gain $a$ and firing rate $y$. 
$\Delta t = 10^{-1}$, $\epsilon _a = \epsilon _b = 10^{-1}$,
$\Gamma = 1$, 
$\lambda _1 = -20$, $\lambda _2 = 18.5$.}
\label{fig_timeSeriesStochasticEscape}
\end{figure}

Increasing the learning rates $\epsilon_a = \epsilon_b$ allows 
the system to escape stochastically from the respective local 
fixed points: The transfer function threshold $b$ conquers the 
local gradient and moves to the other fixed point and back 
(compare Fig.~\ref{fig_stochastic_escape}). 
In the long-term observation the system therefore approximates 
the left and the right part of the target distribution and hence 
minimizes the relative entropy, compare Table~\ref{tab:relativEntropies}. 
These tipping transitions between the two quasi-stationary fixed 
points are illustrated in Fig.~\ref{fig_timeSeriesStochasticEscape}, 
which shows a typical time series for a skewed target distribution.
Note that there are two fixed points for the gain and threshold and 
a direct correspondence to the periods of high and low firing 
rates $y(t)$.

Very low learning rates $\epsilon_a$, $\epsilon_b$, lead to deep 
and big basins of attraction for the respective fixed points, while 
on the other hand high learning rates result in the closely following 
of the threshold to the membrane potential which prohibits reaching 
the target distribution. This mechanism is reminiscent to the
case of Langevin dynamics in a double-well potential
\cite{hanggi1986}, where a stochastically driven particle may 
switch forth and back between two local minima \cite{grosBook}.
The switching time is controlled for the double-well problem by 
the Kramer's escape rate, which depends exponentially on the
potential barrier height. It is difficult to formulate a 
quantitative mapping to the double-well problem, the local
attractors visible in Figs.~\ref{fig_stochastic_escape}
and \ref{fig_timeSeriesStochasticEscape}, and the effective 
barriers in between them, are self-organized structures.
Note that the strength $Q$ of the noise term (\ref{eq_model})
is constant and influences the transition rate only weakly,
due to the continuous adaption of the transfer function,
via (\ref{eq_delta_a}) and (\ref{eq_delta_b}),
to the average strength of the stochastic driving.

\section{Discussion}
We showed that polyhomeostatic adaption of continuous-time leaky integrator
leads to desired firing rate distributions. We also run further simulations
using white noise and Gaussian noise input and replace the transfer function by
other qualitatively different (but still sigmoidal) functions, see Appendix. It
turns out that the polyhomeostatic adaption as well as the self-organized
stochastic escape are quite robust principles. However, the quality of the
approximation (as seen by visual overlapping) and the value of the
Kullback-Leibler divergence depend on the learnings rates, but also on the input
distribution and the input’s strength.

The stochastic tipping as a function of adaption rates has a close
relation to the phenomenon of stochastic escape. The strength of the
driving input noise is constant, but its influence is averaged out
for very low adaption rates. Stochastic escape from one local attractor to
another is not possible. The stochasticity of the input becomes relevant
for intermediate values of adaption rates and stochastic transitions between
the two quasistationary attractors are most frequent. Finally, for very large
adaption rates, the system tips into another dynamical state, tracking
the stochastic input signal nearly instantaneously.
This sequence of behaviors is self organized and reached from any
initial state.



\clearpage \appendix
\section{Polynomial transfer function}
The polyhomeostatic adaption of the system is not changing qualitatively by
replacing the transfer function $g$. Instead it turns out that the system is
robust against changing the transfer function as long as it remains sigmoidal.
We also applied a transfer function
\begin{equation}
g(x) = \frac{\left({x}/{b}\right)^{a b}}
            {\left({x}/{b}\right)^{a b} + 1}~,
\label{eq_g_g0}
\end{equation}
with a polynomial decay to $g(0) = 0$, which limits 
the membrane potential $x \geq 0 $ to be non-negative.
It turns out that the shape of the target distribution $q$ is also well approximated using this
transfer function. Also stochastic escape from one fixed point to another and
back can be observed as for some target distributions two fixed points are
necessary.

\begin{figure}[t]
\centering
\includegraphics[width=0.6\columnwidth]{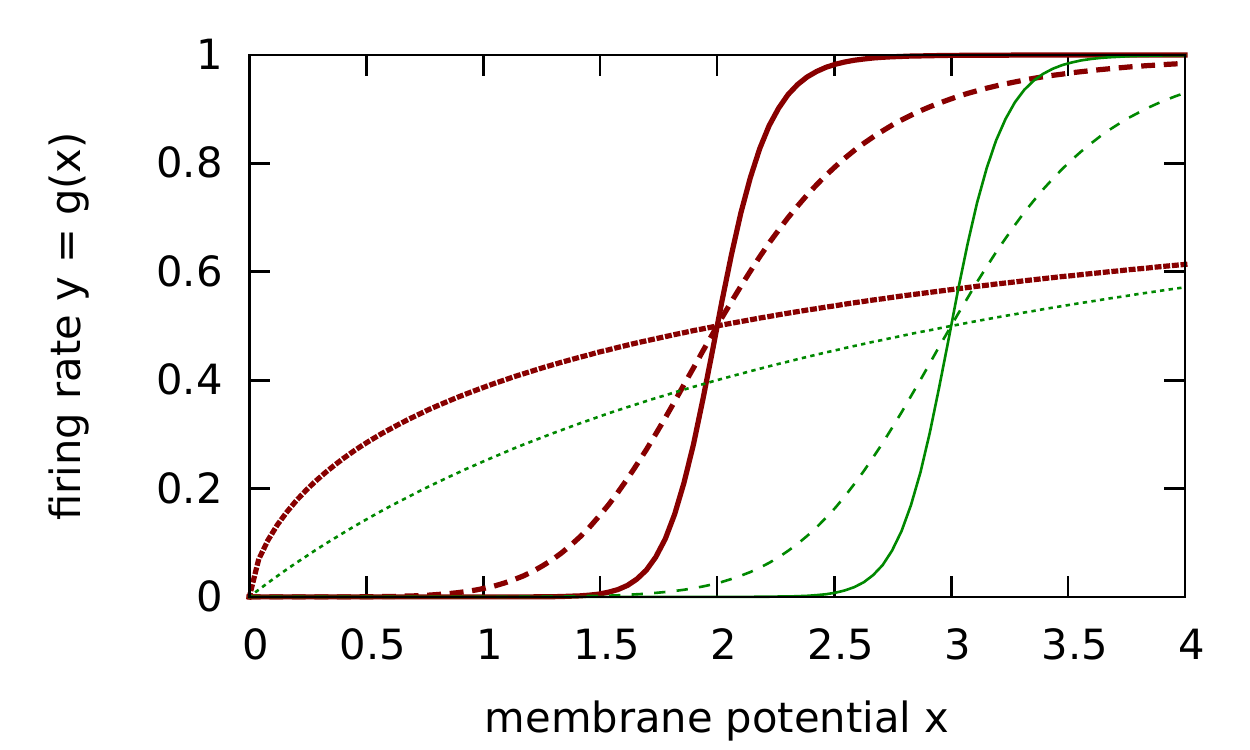}
\caption{The transfer function $g(x)$, see Eq.~(\ref{eq_g_g0}),
for thresholds $b=2$ (red lines) and $b=3$ (green lines) and 
various gains $a$: 1/3 (dotted), 3 (dashed), 9 (solid). No
inflection point is present for exponents $ab<1$.}
\label{fig_transferFunction_g0}
\end{figure}

The transfer
function has an inflection point for
exponents $ab>0$; it is absent for $ab<1$, compare Fig.~\ref{fig_transferFunction_g0}.
The transfer function $g$ behaves as
\begin{equation}
g(x) \ \approx\ \left\{
\begin{array}{cr}
  \left( {x}/{b} \right)^{a b} & x \ll b\\
  \frac{1}{2} + \frac{1}{4} a \left( x - b \right) & x \approx b \\
  1 - \left( {b}/{x} \right)^{a b} & x \gg b
 \end{array}
\right.~.
\end{equation}
The slope is $a/4$ which approaches zero and unity for small and large
membrane potentials respectively.

From Eq.~(\ref{eq_g_g0}) we find the relations
\begin{equation}
\frac{\partial g}{\partial x} = \left(1 - g\right) g \frac{a b}{x}~,
\end{equation}
\begin{equation}
\frac{\partial g}{\partial a} = 
\left(1 - g\right) g b \ln \left(\frac{x}{b}\right),
\qquad
\frac{\partial g}{\partial b} = 
\left(1 - g\right) g a \left[ \ln \left( \frac{x}{b} \right) - 1 \right]~,
\end{equation}
which we can use to evaluate the stochastic adaption rules (\ref{eq_deltaTheta}) as
\begin{equation}
\frac{\mathrm{d}a}{\mathrm{d}t} = \epsilon _a \left[ \frac{1}{a} - b \ln(x/b)  
\Big[ 1 - 2 y + \left( \lambda _1 + 2 \lambda _2 y \right) 
\left( 1 - y \right) y \Big] \right]
\label{eq_delta_a_g0}
\end{equation}
and
\begin{equation}
\frac{\mathrm{d}b}{\mathrm{d}t} = \epsilon _b \left[ \frac{1}{b} - a \Big[ 
\ln(x/b) - 1 \Big]\,
\Big[ 1 - 2 y + \left( \lambda _1 + 2 \lambda _2 y \right) 
\left( 1 - y \right) y \Big] \right]~.
\label{eq_delta_b_g0}
\end{equation}

\begin{table}[b]
	\caption{\label{tab:relativEntropiesVariousTargetDistributions_g0}Relative entropies of various target distributions compared to the corresponding achieved distribution ($\epsilon _a = \epsilon _b = 10^{-2}$, $\mathrm{bins} = 100$).}
\begin{indented}
	\lineup
\item[]\begin{tabular}{llll}
	\br
	$\lambda _1$ & $\lambda _2$ & shape & $D_{KL}$ \tabularnewline
	\mr
	0 & 0 & uniform & 0.060131 \tabularnewline
	-10 & 0 & left dominant & 0.069351 \tabularnewline
	+10 & 0 & right dominant & 0.114578 \tabularnewline
	-10 & +10 & left/right dominant & 0.051811 \tabularnewline
	\mr
	+20 & -20 & hill & 0.148098 \tabularnewline
	-20 & +20 & left/right, symmetric & 0.189217 \tabularnewline
	-20 & +19 & left/right, left skewed & 0.063934 \tabularnewline
	-20 & +18.5 & left/right, left skewed & 0.261215 \tabularnewline
	\br
\end{tabular}
\end{indented}
\end{table}

Applying this transfer function $g$ it turns out that the target distribution
is well approximated also in this case, even though the membrane potential is restricted to
non-negative numbers. Tab.~\ref{tab:relativEntropiesVariousTargetDistributions_g0}
lists the well minimized Kullback-Leibler divergences for several target distributions.

We conclude that the stochastic adaption rules are therefore generic and qualitatively
independent on the concrete realization of the transfer function. However, quantitatively the
resulting relative entropies depend on the choice of the transfer function which also
influences the optimal adaption rates $\epsilon_a$ and $\epsilon_b$.

%
\subsection{Self-Organized Stochastic Escape}

\begin{figure}[t]
\centering
\includegraphics[width=0.59\columnwidth]{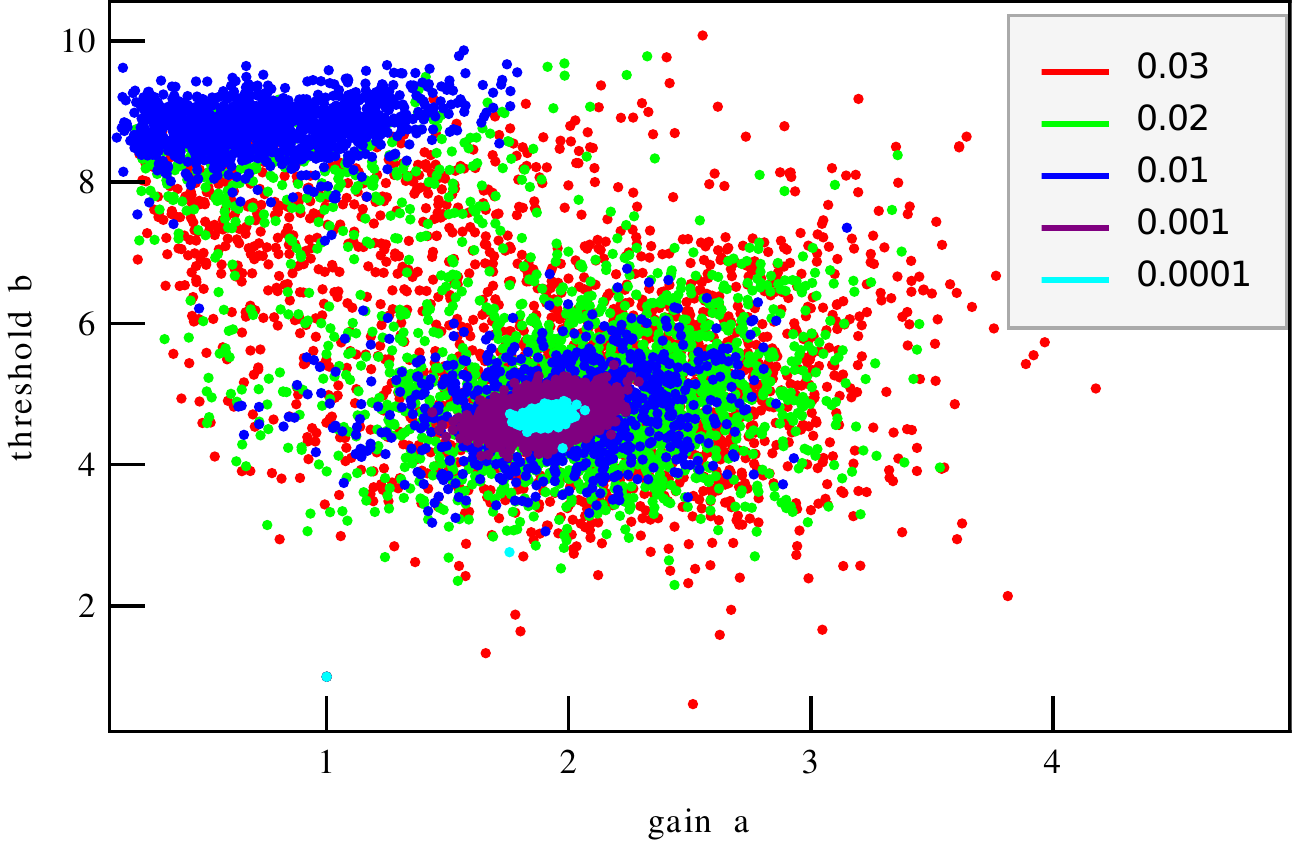}
\includegraphics[width=0.4\columnwidth]{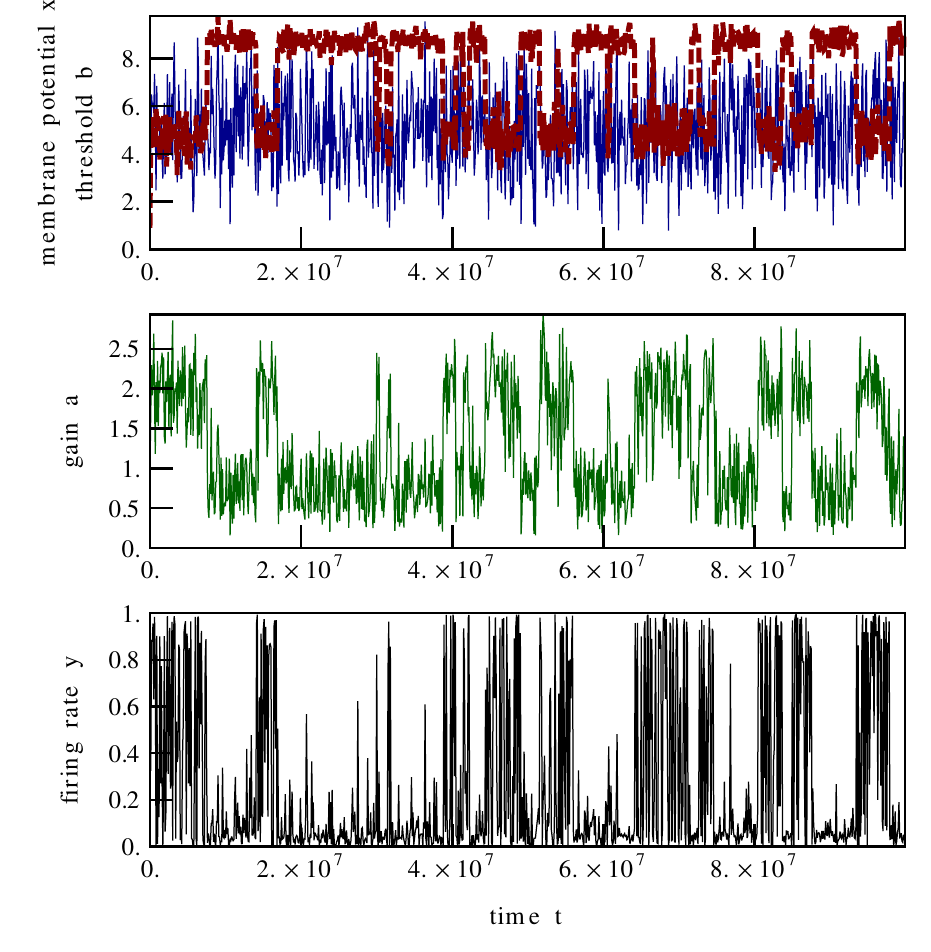}
\caption{Left: Stochastic escape: Phase diagram of the transfer 
function gain vs.\ transfer function threshold for a convex 
left skewed target distribution with various learning 
rates ($\epsilon _a$ and $\epsilon _b$ given in the legend).
$\Delta t = 10^{-2}$, $\Gamma  = 0.1$, 
$\lambda _1 = -20$, $\lambda _2 = 19$.
Right: Time series: membrane potential $x$, 
transfer function threshold $b$ (dashed), 
transfer function gain $a$ and firing rate $y$. 
$\Delta t = 10^{-1}$, $\epsilon _a = \epsilon _b = 10^{-2}$, 
$\Gamma  = 0.1$, 
$\lambda _1 = -20$, $\lambda _2 = 19$.}
\label{fig_stochastic_escape_Appendix}
\end{figure}

For the non-symmetric convex target distribution 
($\lambda _1 = -20$, $\lambda _2 = 19$) there are two fixed 
points. Since the target distribution cannot be well approximated 
by only one fixed point the system escapes stochastically from
one to the other and back with a certain period, compare
Fig.~\ref{fig_stochastic_escape_Appendix}.
For small learning rates $\epsilon _a =
\epsilon _b \lessapprox 0.01$ the system in trapped in only one 
fixed point. The relative entropy therefore is not well minimized.

\begin{table}[b]
\caption{\label{tab:relativEntropies_g0}Relative entropies of the left-skewed target distribution ($\lambda _1 = -20$, $\lambda _2 = 19$) compared to the achieved distribution for various learning rates $\epsilon _a$ and $\epsilon _b$. Note that the Kullback-Leibler divergence
is not minimized for $\epsilon_b \gtrapprox 0.05$ due to the fast correlation of the
membrane potential and the transfer function threshold.}
\begin{indented}
	\lineup
\item[]\begin{tabular}{l|lllllll}
	\br
	$\epsilon _a = \epsilon _b$ & $10^{-4}$ & $10^{-3}$ & 0.01 & 0.03 & 0.04 & 0.05 & 0.06 \tabularnewline
	\mr
	$D_{KL}$ & 0.376 & 0.368 & 0.064 & 0.043 & 0.017 & 1.892 & 1.591 \tabularnewline
	\br
\end{tabular}
\end{indented}
\end{table}

For intermediate learning rates 
$0.01 \lessapprox \epsilon _a = \epsilon _b \lessapprox 0.04$ 
the perturbation is high enough to stochastically escape from
that fixed point and approach another one.
Fig.~\ref{fig_stochastic_escape_Appendix}
shows a typical time series for this tipping. This has also 
an effect on the relative entropy which therefore is even smaller 
than without stochastic escape
(see Tab.~\ref{tab:relativEntropies_g0}).

For high learning rates $\epsilon _a = \epsilon _b \gtrapprox 0.05$ 
the system’s behavior changes: the transfer function is close to a 
Heaviside step function and the threshold follows the membrane 
potential quickly. In that state the achieved distribution is not 
close to the target distribution, therefore the
relative entropy is not minimized anymore (see
Tab.~\ref{tab:relativEntropies_g0}).

\end{document}